\documentclass[aps,prfluids,preprint,amsmath,amssymb,floatfix]{revtex4-2}

\makeatletter
\def\pre@bibdata{PRFMainNotes}
\makeatother

\usepackage{graphicx}%
\usepackage{multirow}%
\usepackage{amsmath,amssymb,amsfonts}%
\usepackage{mathrsfs}%
\usepackage{xcolor}%
\usepackage{textcomp}%
\usepackage{booktabs}%
\usepackage{siunitx}
\usepackage{hyperref}
\hypersetup{colorlinks=true, linkcolor=blue, citecolor=blue, urlcolor=blue}

\begin{document}
\title{Two-way coupling in active suspensions suppresses mobility-selective accumulation and induces non-monotonic flow stabilization}

\author{Ziyue Yu}
\email{ZIY79@pitt.edu}
\affiliation{Department of Civil and Environmental Engineering, University of Pittsburgh, Pittsburgh, Pennsylvania 15261, USA}
\author{Aineo M. Wu}
\affiliation{Simon's Rock at Bard College, Barrytown, New York 12507, USA}
\author{Lei Fang}
\email{lei.fang@pitt.edu}
\affiliation{Department of Civil and Environmental Engineering, University of Pittsburgh, Pittsburgh, Pennsylvania 15261, USA}
\affiliation{Department of Mechanical Engineering and Materials Science, University of Pittsburgh, Pittsburgh, Pennsylvania 15261, USA}
\affiliation{Department of Bioengineering, University of Pittsburgh, Pittsburgh, Pennsylvania 15261, USA}

\begin{abstract}
Swimmers live in moving fluids: the flow organizes their spatial distribution, and their collective motion reshapes the flow in turn. We experimentally study the two-way coupling between a swarm of centimeter-scale swimmers
(\textit{Artemia salina}) and an electromagnetically driven quasi-two-dimensional cellular flow. We report two key findings, one in each direction of the coupling. First, at low swimmer density, the swarm recovers the mobility-selective accumulation onto the attracting hyperbolic Lagrangian coherent structures (LCSs) predicted by one-way-coupled simulations: the accumulation is strongest when the swimming speed is comparable to the flow speed. As the density increases, the swarm still accumulates on the LCSs, but the mobility selectivity is progressively lost. Second, we examine the back-action of the swarm on the LCSs. A swarm has previously been found to disrupt the LCSs. We find that the response depends on the swimmer density and the background forcing. Denser swarms tend to disrupt the LCSs more strongly, as previously reported. However, at stronger forcing, adding a moderate number of swimmers instead stabilizes the LCSs.
\end{abstract}

\maketitle

\section{Introduction}

Active matter, systems whose constituents consume stored or ambient energy to propel themselves, spans a wide range of scales, from microscopic bacteria and algae to centimeter-scale crustaceans and meter-scale schooling animals \cite{ramaswamy2010,marchetti2013,bechinger2016}. Such swimmers rarely inhabit quiescent fluid: oceans, lakes, and other natural waters are continuously stirred into spatially structured, time-dependent flows, within which organisms must move, feed, and aggregate \cite{durham2013,guasto2012}. The interaction between active matter and such structured flows is therefore central to the transport and mixing that occur in them.

In one direction of this interaction, the flow organizes the spatial distribution of the swimmers. The simplest description treats each swimmer as an inertialess, non-spherical active particle advected by the flow and neglects its back-action on the flow; this is the one-way-coupling approximation \cite{si2021,si2022,khurana2011}. Two features distinguish active particles from passive point tracers: an elongated shape, so that their orientation is set by the local strain rate and vorticity, and intrinsic mobility, so that their trajectories depart from those of fluid elements. Their spatial distribution is consequently much more heterogeneous than that of passive point tracers. Depending on the flow and on their shape and mobility, swimmers may be trapped in some regions of the flow and depleted from others, or may accumulate along particular flow structures. In chaotic flows, hydrodynamic trapping within the recirculating regions can reduce the large-scale transport of swimmers \cite{khurana2011,khurana2012,qin2022}. Even a simple shear flow produces such heterogeneity: motile bacteria are trapped in the high-shear regions of a channel and depleted from the low-shear ones \cite{rusconi2014}. In turbulence, motile phytoplankton form patches on scales much smaller than the integral scale \cite{durham2013}. Elongated swimmers also align preferentially with the local flow \cite{borgnino2019}. Si and Fang \cite{si2021} showed that such swimmers align with and accumulate along the attracting hyperbolic Lagrangian coherent structures (LCSs) of the flow. The accumulation depends on the aspect ratio and on the swimmer's mobility. These studies have largely focused on one-way coupling, in which the back-action of the swimmers on the flow is neglected.

The back-action from collective swimmers to the flow is not always negligible. Swimmers inject energy into the surrounding fluid, and this biogenic agitation can contribute to turbulence and mixing \cite{huntley2004,kunze2006,katija2009}. In quiescent fluid, this agitation generates flow across a range of scales. The diffusion of passive tracers is enhanced by swimmer agitation \cite{wu2000,leptos2009,thiffeault2010}. Dense bacterial suspensions sustain turbulence-like flows of their own \cite{wensink2012,dunkel2013}. The collective vertical migration of zooplankton generates aggregation-scale jets \cite{wilhelmus2014,mohebbi2024} that mix a stratified water column \cite{houghton2018,ouillon2020}. When a background flow is already present, swimmers modify its structure instead. Swimming bacteria lower the effective viscosity of a sheared suspension \cite{sokolov2009,lopez2015}. A swarm deforms the attracting LCSs of a background flow, and the deformation is much stronger than that produced by random agitation of the same energy \cite{si2024}. For a single swimmer in a background shear, the biologically generated energy flux can either intensify or attenuate the shear, depending on the geometry between the agitation and the shear \cite{si2024tensor}. Bacteria that accumulate along the boundaries of Lagrangian vortices strengthen the barriers those boundaries pose to scalar transport \cite{ran2024}.

The two directions of the coupling are not independent: the flow structure, together with the swimmers' shape and mobility, determines the swimmer distribution, and the swimmers alter that structure by injecting energy into the fluid. They have nonetheless usually been studied separately, and the two-way coupling between swimmers and the flow structure has received little systematic attention. Treating the coupling as two-way raises two questions: whether the background flow still organizes the swimmer distribution as the one-way-coupled approximation predicts, and how the swimmers modify that flow in return.

In this work, we study the two-way coupling experimentally by incorporating a swarm of the centimeter-scale swimmer \textit{Artemia salina} into an electromagnetically driven quasi-two-dimensional cellular flow, spanning a wide range of swimmer counts $N$ and background forcing $E$. We characterize the flow structure by its attracting hyperbolic LCSs, which form the skeleton of the flow. We report two key findings. First, a swarm of finite-sized swimmers recovers the mobility-selective accumulation predicted by one-way-coupled simulations of inertialess swimmers \cite{si2021}: at the fixed aspect ratio of \textit{A. salina}, the accumulation is strongest when the swimming speed is comparable to the flow speed. This agreement holds only at low swimmer concentrations, where the swarm does not appreciably modify the background flow. As the swimmer density increases the coupling becomes two-way, and the mobility selectivity is progressively lost, although the net accumulation on the LCSs persists.

Second, the swarm can either disrupt or stabilize the LCSs, depending on the swimmer density and the background forcing. A dense swarm in a weakly forced flow disrupts the LCSs, as reported previously \cite{si2024}. However, at stronger forcing, where the background flow becomes weakly time-dependent, the presence of a swarm of intermediate density stabilizes the LCSs.

\section{Experiment: electromagnetically driven\texorpdfstring{\\}{ }quasi-two-dimensional cellular flow}\label{sec:exp_intro}
\subsection{Apparatus}

\begin{figure*}
\centering
\includegraphics[width=1\linewidth]{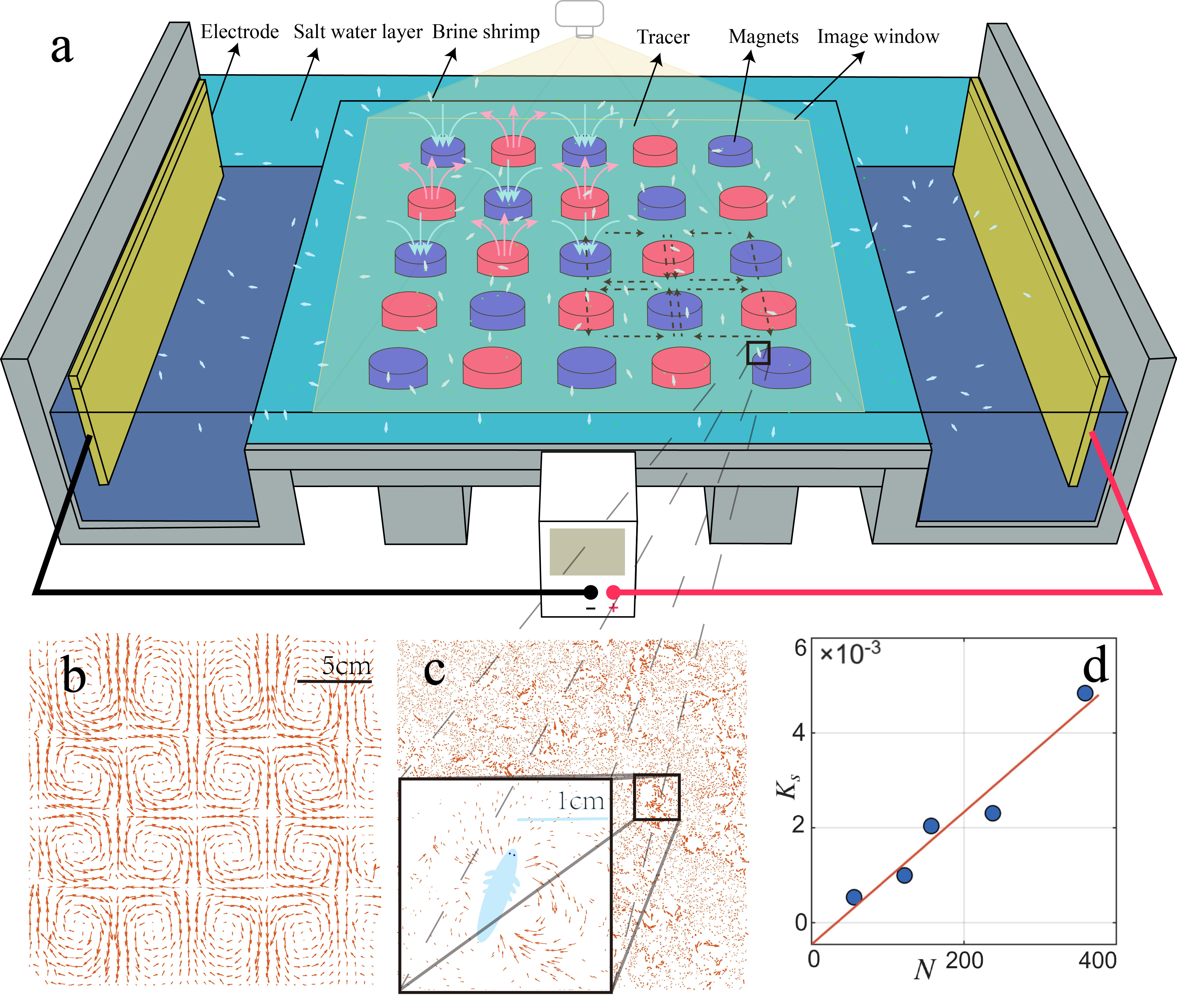}
\caption{Experimental setup and representative flow fields. \textbf{(a)} Schematic of the electromagnetically driven thin-layer apparatus (not to scale). Blue and red magnets denote field lines pointing into and out of the plane. The yellow-shaded region marks the $20\times20~\mathrm{cm}^2$ imaging window. \textbf{(b)} Background velocity field without shrimp, showing the vortex lattice imposed by the magnets (scale bar $5~\mathrm{cm}$). \textbf{(c)}
Instantaneous field with shrimp (images of a dilute and a dense run are shown in Movie~S1 of the Supplemental Material). The inset enlarges the boxed region around a single \textit{A.~salina}, showing its recirculating disturbance (scale bar $1~\mathrm{cm}$). \textbf{(d)} Mean kinetic energy $K_s$ of the shrimp-driven flow measured in five calibration runs (no background current) as a function of the number of swimmers $N$. The solid line is a linear least-squares fit.}
\label{fig:setup}
\end{figure*}

The background flow is a quasi-two-dimensional (quasi-2D) cellular flow produced by an electromagnetically driven thin-layer system (Fig.~\ref{fig:setup}(a)). The apparatus has been described in detail previously \cite{kelley2011,fang2018boundaries,si2022,si2024,yu2026tensor}; here we restate only the features relevant to the present measurements. The working fluid is a uniform $6~\mathrm{mm}$ layer of saltwater ($14\%$ NaCl by mass) resting on a tempered-glass substrate above a checkerboard array of permanent magnets of center-to-center spacing $L_m=5.1~\mathrm{cm}$. A direct current driven between a pair of copper electrodes interacts with the magnetic field to produce Lorentz body forces that reverse in sign from site to site and drive a square array of counter-rotating vortices of size $L_m$ (Fig.~\ref{fig:setup}(a)). Below a critical forcing the flow in such a layer remains essentially two-dimensional, a condition satisfied throughout the present experiments \cite{kelley2011}. The flow field, spanning $20\times20~\mathrm{cm}^2$, was recorded at 60 frames per second with an industrial camera and analyzed by particle tracking velocimetry (PTV), as described in Sec.~\ref{sec:ptv}. Full details of the fluid properties, magnet specifications, and surface treatment are given in the Supplemental Material, Sec.~S1.1.

\subsection{Particle tracking and data post-processing}\label{sec:ptv}

The flow was seeded with green-fluorescent polyethylene tracers (Cospheric; density $1.0125~\mathrm{g\,cm^{-3}}$, diameter $106$--$125~\mathrm{\mu m}$) and imaged from above. Being slightly less dense than the working fluid, the tracers remain at the free surface and follow the quasi-2D surface flow; the Stokes number is below $2\times10^{-4}$ throughout, so they track the fluid motion faithfully \cite{ouellette2008transport}. A small amount of surfactant was added to prevent surface-tension-driven clustering, and the absence of tracer motion in the unforced layer confirms that residual surface-tension effects are negligible. Particle positions, velocities, and accelerations were obtained with the PTV algorithm of Ref.~\cite{ouellette2006ptv}; camera and seeding specifications are given in the Supplemental Material, Sec.~S1.4.

When shrimp (Sec.~\ref{sec:shrimp}) were present in the field of view, the pixels belonging to each animal were masked to the background intensity before the images were passed to the PTV algorithm, so that only the fluid tracers were tracked and the recovered velocity field does not contain the motion of the animals themselves. The scattered velocity vectors were interpolated onto a regular Eulerian grid of spacing $\approx1.25~\mathrm{mm}$ for the field-based diagnostics of Sec.~\ref{sec:ftle}.

\subsection{Flow characterization}\label{sec:flowchar}

The background flow is characterized by its root-mean-square velocity $U=\sqrt{\langle\mathbf{u}\cdot\mathbf{u}\rangle}$, where $\mathbf{u}$ is the velocity field measured in that condition and $\langle\cdot\rangle$ denotes a space--time average. We tune the forcing by varying the dc current from $0.1$ to $2.3~\mathrm{A}$, which raises $U$ from $0.1$ to $1.75~\mathrm{cm\,s^{-1}}$. With the magnet spacing $L_m$ as the characteristic length, the Reynolds number $Re=UL_m/\nu$ ranges from $40$ to $700$. At the lowest forcing the flow is a steady cellular array, becoming progressively more time dependent as the current increases. At higher forcing the cellular pattern is no longer steady and the flow is weakly turbulent. The strength of the forcing is quantified by the mean kinetic energy per unit mass,
\begin{equation}\label{eq:E}
E=\tfrac{1}{2}\langle\mathbf{u}\cdot\mathbf{u}\rangle=U^2/2,
\end{equation}
and the eddy turnover time $T_e=L_m/U$ sets the natural time scale.

\section{Experiment: swarming active matter}\label{sec:shrimp}

The swimmers are adult brine shrimp, \textit{Artemia salina}, with a body length of approximately $1~\mathrm{cm}$ and an aspect ratio of around $3$, defined as the ratio of body length to the maximum body width across the appendages. They swim by beating their appendages, generating a propulsive jet directed opposite to the swimming direction (Fig.~\ref{fig:setup}(c), inset). The animals were obtained from a commercial supplier (Northeast Brine Shrimp) and held in saltwater ($3\%$ NaCl by mass) for at least $24$~h before use.
Because the animals are less dense than the working fluid, they float at the free surface and remain in the quasi-2D measurement plane throughout an experiment. The mean single-animal swimming speed, averaged over the five quiescent runs (one at each density, with no background flow), is $V_s = 0.54~\mathrm{cm\,s^{-1}}$ with a run-to-run standard deviation of $0.06~\mathrm{cm\,s^{-1}}$ (Supplemental Material, Fig.~S2(a)). Because $V_s$ is essentially independent of swarm density, it is treated as a constant.

\subsection{Protocol for the shrimp experiments}\label{sec:protocol}

Experiments proceeded in two stages: the shrimp-free background flow was characterized first, and the forcing sweep was then repeated at successively higher swimmer densities. Data were collected according to the following protocol.

\begin{enumerate}
\item[(i)] Load the working fluid into the apparatus.
\item[(ii)] With no shrimp present, set each dc current, wait until the flow reaches a statistically stationary state, and record the background flow.
\item[(iii)] Introduce \textit{A. salina} into the working fluid and wait until the animals spread uniformly across the imaging domain and the swarm reaches a statistically stationary state.
\item[(iv)] With the dc power supply off, record the flow driven by the swarm alone.
\item[(v)] Repeat the current sweep of step (ii) with the swarm present.
\item[(vi)] Turn off the dc power supply, add more shrimp, and repeat steps (iii)--(v), so that the density increases monotonically over five densities.
\end{enumerate}

The densities and the current levels used at each density are listed in Table~S1 of the Supplemental Material.

\subsection{Characterization of the swarm-induced flow}\label{sec:swarmflow}

Under the spatially uniform illumination, the animals exhibit no directional bias: the shrimp-induced velocity field is statistically isotropic at every swimmer density, consistent with our previous experiments \cite{si2024}. The group is therefore in a swarming state \cite{si2024}. The distributions supporting this isotropy are given in the Supplemental Material, Fig.~S2(b).

The flow generated by the swarm is characterized by the mean kinetic energy per unit mass
\begin{equation}\label{eq:Ks}
K_s=\tfrac{1}{2}\langle\mathbf{u}_s\cdot\mathbf{u}_s\rangle,
\end{equation}
where $\mathbf{u}_s$ is the field generated by swimmers in quiescent fluid. Measured over the entire swarm, $K_s$ is the kinetic energy maintained in the layer by the swarm, and it grows linearly with the number of swimmers $N$ (Fig.~\ref{fig:setup}(d)).

\section{Theoretical background}\label{sec:theory}

\subsection{Two-dimensional Navier--Stokes equation and energy balance}\label{sec:theory_ns}

In the thin-layer flow, the measured velocity field is well described by the two-dimensional Navier--Stokes equation:
\begin{equation}\label{eq:ns}
\partial_t\mathbf{u}+(\mathbf{u}\cdot\nabla)\mathbf{u} =-\frac{1}{\rho}\nabla p+\nu\nabla^{2}\mathbf{u}-\alpha\mathbf{u}+\mathbf{f},
\end{equation}
where $\mathbf{f}$ is the Lorentz body force and the term $-\alpha\mathbf{u}$ models the friction exerted by the substrate on the fluid layer \cite{boffetta2012,fang2017decay}. At the statistically stationary state reached before every acquisition, the power injected by any source is therefore dissipated by the same friction. For the electromagnetically driven flow, the balance is $P_\mathrm{EM}=\alpha\langle\mathbf{u}\cdot\mathbf{u}\rangle=2\alpha E$, where $P_\mathrm{EM}$ is the power from the electromagnetic force. The flow driven by the swarm alone satisfies the same balance, $P_\mathrm{shrimp}=2\alpha K_s$, where $P_\mathrm{shrimp}$ is the power from the shrimp swarm.

Here we define the power ratio:
\begin{equation}\label{eq:powerratio}
\frac{P_\mathrm{shrimp}}{P_\mathrm{EM}}=\frac{2\alpha K_s}{2\alpha E}=\frac{K_s}{E}.
\end{equation}
It quantifies the strength of the swarm's forcing power relative to the electromagnetic forcing power.

The flow-swimmer interaction involves two dimensionless ratios: the velocity ratio $V_s/U$, comparing the single-swimmer speed with the background flow speed, and the power ratio $K_s/E$ of Eq.~(\ref{eq:powerratio}). A given $K_s/E$ can be reached either at fixed $N$ by changing the forcing or at fixed $E$ by changing $N$, so that cases with very different swimmer densities may share the same value, which would hide the dependence on swarm size. We therefore parametrize the parameter space with $V_s/U$ and the density-weighted ratio $NK_s/E$; the extra factor of $N$ separates the two routes and makes the dependence on swimmer density explicit.

\subsection{Attracting hyperbolic Lagrangian coherent structures and the finite-time Lyapunov exponent}\label{sec:ftle}

We analyze the flow in a Lagrangian frame and detect coherent structures from the flow map defined below. All Lagrangian fields are built from the flow map
\begin{equation}\label{eq:flowmap}
\mathbf{F}_{t_0}^{t}(\mathbf{x}_0)=\mathbf{x}(t;t_0,\mathbf{x}_0), \qquad \mathbf{x}(t;t_0,\mathbf{x}_0)=\mathbf{x}_0+\int_{t_0}^{t}\mathbf{u}(\tau,\mathbf{x}(\tau;t_0,\mathbf{x}_0))\,\mathrm{d}\tau .
\end{equation}
Hyperbolic LCSs are the material lines that attract or repel neighboring fluid most strongly, and they act as the transport barriers organizing the cellular pattern \cite{haller2015}. Throughout this work, we focus on the attracting branch since previous work has shown the swimmers tend to align with attracting LCSs \cite{si2021}. We detect them with the finite-time Lyapunov exponent \cite{shadden2005}: with $\lambda_2$ the largest eigenvalue of the right Cauchy--Green strain tensor $\mathbf{C}=(\nabla\mathbf{F}_{t_0}^{t})^{\mathsf{T}}\nabla\mathbf{F}_{t_0}^{t}$,
\begin{equation}\label{eq:ftle}
\mathrm{FTLE}(\mathbf{x}_0)=\frac{1}{t-t_0}\,\tfrac{1}{2}\ln\lambda_2(\mathbf{x}_0).
\end{equation}
Ridges of the backward-time field ($t<t_0$) mark the attracting LCSs ($\mathrm{FTLE}^{-}$). The flow map is integrated over one eddy turnover time $T_e$.   The $\mathrm{FTLE}^{-}$ ridges are well-established for identifying LCSs. Numerical details of the flow-map integration and of the differentiation scheme are given in the Supplemental Material, Sec.~S2.1.

\section{The impact of the flow on the swarm: mobility-selective accumulation and the density-dependent loss of its selectivity}\label{sec:results_acc}

A prior numerical study established that non-spherical swimmers in a background flow align with, and accumulate onto, the attracting LCSs \citep{si2021}. The prior study used a hybrid approach: inertialess rod-like swimmers were integrated based on experimentally measured velocity fields and reoriented by the local velocity gradients through Jeffery's equation \citep{jeffery1922}, under the assumption of one-way coupling, in which the swimmers do not themselves modify the background flow field. Under the one-way coupling assumption, the strength of the resulting accumulation depends on the swimmer mobility: the accumulation is most pronounced when the swimming speed is comparable to the flow speed.

To test whether the conclusion holds under a two-way coupling scenario, we measure the shrimp number density inside the attracting-LCS region relative to that over the whole domain. We identify the attracting-LCS region as the set of grid points whose $\mathrm{FTLE}^{-}$ value lies in the lowest $15\%$ of the field.
The choice of threshold is not critical: cuts of $10\%$, $15\%$, and $20\%$ yield the same behavior (Supplemental Material, Sec.~S2.3). We use brine shrimp centroids to define their location. We define the accumulation coefficient as:
\begin{equation}\label{eq:cc}
CC=\frac{N_{\mathrm{LCS}}/A_{\mathrm{LCS}}}{N_{\mathrm{tot}}/A_{\mathrm{tot}}},
\end{equation}
where $N_{\mathrm{LCS}}$ is the number of shrimp inside the LCS region of area $A_{\mathrm{LCS}}$, and $N_{\mathrm{tot}}$, $A_{\mathrm{tot}}$ are the corresponding totals over the interrogation area. By construction, $CC=1$ indicates a uniform distribution, $CC>1$ means preferential accumulation on the attracting LCSs, and $CC<1$ means depletion.

\begin{figure*}[t]
\centering
\includegraphics[width=1\linewidth]{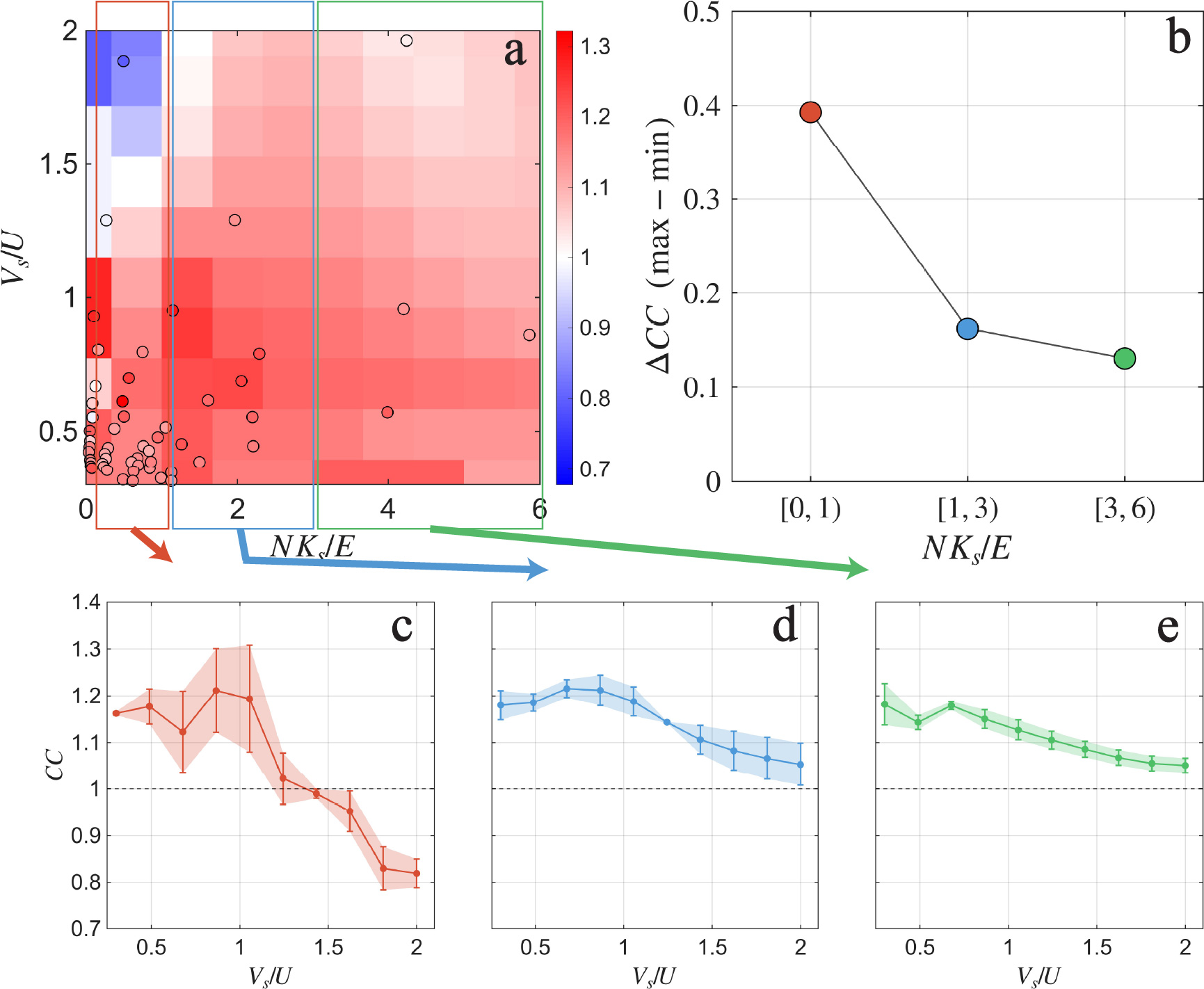}
\caption{Accumulation effect as a function of the control parameters $NK_s/E$ and $V_s/U$. \textbf{(a)} Heat map of the accumulation coefficient $CC$ (Eq.~\ref{eq:cc}) in the $(NK_s/E,\,V_s/U)$ plane. Circles mark the measured experimental conditions, colored by their measured $CC$; the surrounding field is obtained by interpolation. Red color denotes stronger accumulation onto the attracting LCSs ($CC>1$); blue color indicates depletion ($CC<1$). The red, blue, and green boxes outline the three $NK_s/E$ ranges examined in (c--e). \textbf{(b)} Accumulation range $\Delta CC=CC_{\max}-CC_{\min}$, taken over $V_s/U$ within each $NK_s/E$ bin, versus $NK_s/E$. \textbf{(c--e)} $CC$ versus $V_s/U$ for the three bins outlined in (a): \textbf{(c)} $NK_s/E\in[0,1]$, \textbf{(d)} $NK_s/E\in[1,3]$, and \textbf{(e)} $NK_s/E\in[3,6]$. Markers and error bars give the mean and standard deviation within each bin; shaded bands indicate the spread, and the dashed line marks $CC=1$.}
\label{fig:accumulation}
\end{figure*}

Our experiments reproduce the accumulation behavior and confirm that it holds in a real swarm, as long as the swimmers' density remains small. Figure~\ref{fig:accumulation}(a) maps $CC$ across the $(NK_s/E,\,V_s/U)$ plane. In the low-$NK_s/E$ region (red box) the shrimp accumulate on the attracting LCSs ($CC>1$) when $V_s/U$ is near one --- so that the accumulation is strongest when the swimming speed is comparable to the background flow speed. This optimum is consistent with a competition between mobility and the flow's reorientation ability: swimmers that are too slow are reoriented by the flow and rarely reach the attracting ridges, while swimmers that are too fast leave the reorientation regions before they are captured \cite{si2021}.

This dilute-limit behavior changes once the swarm is dense enough to perturb the background flow appreciably. Within each density bin, we define the accumulation range $\Delta CC = CC_{\max}-CC_{\min}$, the difference between the largest and smallest values of $CC$ over $V_s/U$ (Fig.~\ref{fig:accumulation}(b)). $\Delta CC$ decreases monotonically with $NK_s/E$, from $\approx0.4$ in the most dilute bin to around 0.13 at the highest density. As the density increases, the peak of $CC$ near $V_s/U$ around 1 in the most dilute bin flattens, and
$CC$ becomes nearly independent of $V_s/U$ (Fig.~\ref{fig:accumulation}(c--e)). What is removed is the mobility selectivity rather than the accumulation itself: Depletion ($CC<1$) is observed only at the most dilute loading, for $V_s/U>1.2$; at every other loading, the swarm accumulates on the attracting LCSs ($CC>1$) under all measured conditions. The dependence of $CC$ on $V_s/U$ flattens from the peaked form of the dilute limit (Fig.~\ref{fig:accumulation}(c)) to a nearly constant level (Fig.~\ref{fig:accumulation}(d,e)).

One possible interpretation is that the loss of mobility selectivity at higher swimmer density arises from the increasing back-action of the swimmers on the flow. As the swarm more strongly modifies the attracting LCSs—either stabilizing or degrading them, as discussed in the next section—the swimmer distribution and the LCS organization become increasingly coupled. This feedback may weaken the direct dependence of accumulation on the mobility ratio \(V_s/U\), while maintaining a persistent preference for the attracting LCSs. The detailed mechanism underlying this apparent self-regulation remains unclear and deserves further mechanistic investigation.

\section{The impact of the swarm on the flow: regime-dependent disruption and stabilization of the flow skeleton}\label{sec:results_stab}

We now examine the swimmer swarm's impact on the flow. Previous work showed that this effect increases monotonically with swimmer density, i.e., as the swimmer density increases, the attracting LCSs become progressively more distorted, and the structural integrity of the LCSs deteriorates \cite{si2024}. However, over the broader parameter range explored here, we find that the swimmer swarm’s impact on LCS integrity is non-monotonic.

Before discussing these behaviors, we distinguish two terms used to characterize changes in the LCSs. Degradation refers to the broadening or smearing of an LCS ridge, whereas dislodgement refers to the displacement of the ridge away from the underlying cellular lattice. In the latter case, increasing time dependence causes the LCSs to wander, progressively departing from the cellular organization imposed by the external forcing.

\begin{figure*}[!t]
\centering
\includegraphics[width=1\linewidth]{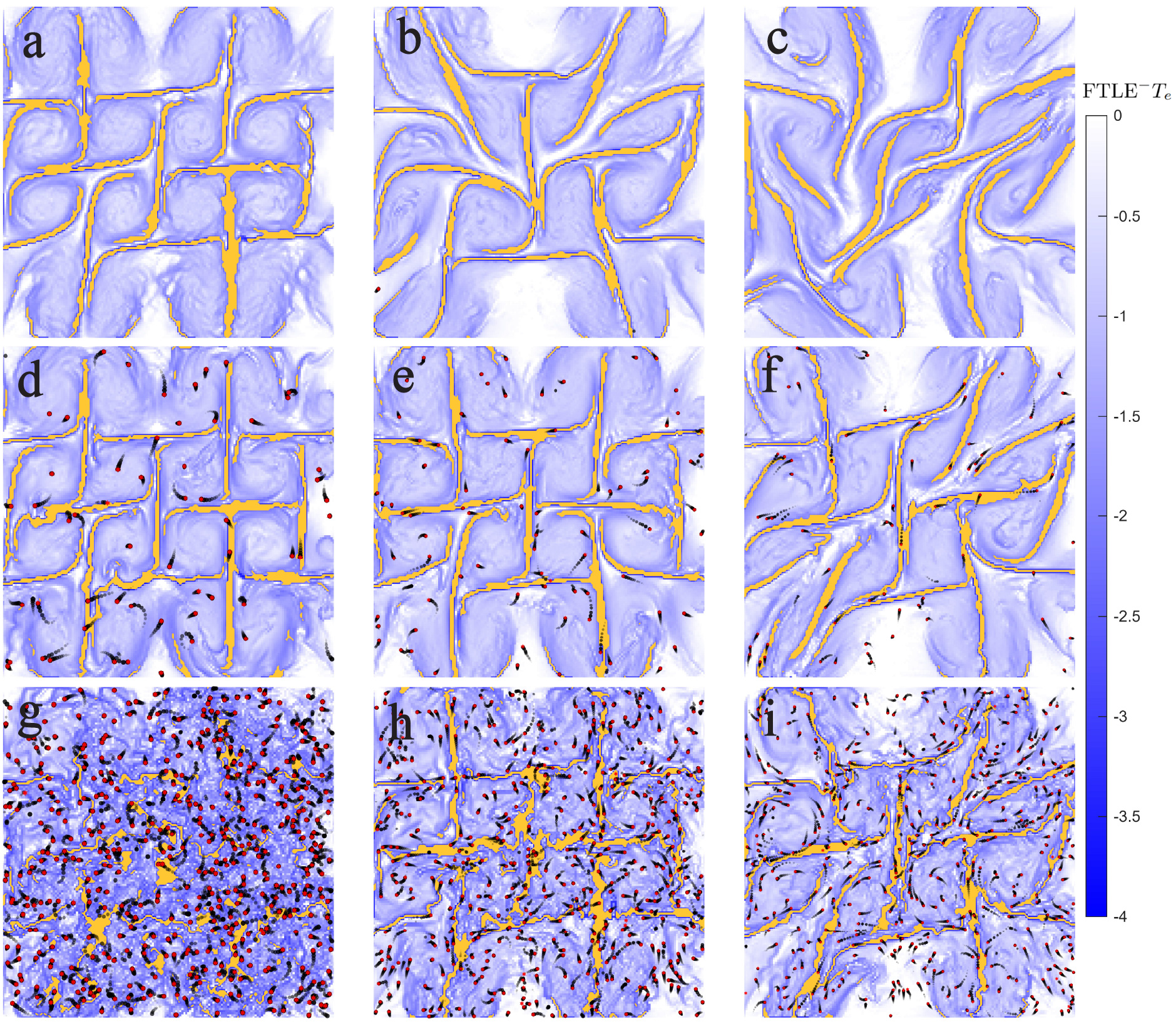}
\caption{Attracting LCSs under increasing background forcing $E$ and swimmer count $N$. In each panel, the blue color shows the dimensionless $\mathrm{FTLE}^{-}$ (the backward FTLE is normalized by the eddy turnover time), and the yellow region marks its lowest $15\%$, i.e.\ the most strongly attracting ridges (cf.\ Sec.~\ref{sec:results_acc}). Columns correspond to increasing $E$ from left to right; rows correspond to increasing $N$ from top to bottom. All values of $E$ are in $\mathrm{cm^2\,s^{-2}}$. \textbf{(a--c)} Shrimp-free background flow ($N=0$) at $E=0.06$, $0.30$, and $0.82$. \textbf{(d--f)} Low swimmer density ($N\approx60$) at $E=0.04$, $0.24$, and $0.79$. \textbf{(g--i)} High swimmer density ($N\approx410$) at $E=0.02$, $0.26$, and $0.77$. Red dots mark the shrimp centroids and the short black trails mark their recent trajectories over the past $0.1$--$0.2T_e$. }

\label{fig:ftle}
\end{figure*}

Figure~\ref{fig:ftle} shows the attracting LCSs across the $(N,E)$ parameter space, with $E$ increasing from left to right and $N$ from top to bottom; Movie~S2 of the Supplemental Material shows the same nine conditions in time. In the shrimp-free flow (Fig.~\ref{fig:ftle}(a--c)), the LCSs become progressively dislodged and disorganized as $E$ increases, reflecting the breakup of the cellular pattern and the onset of time dependence. These shrimp-free cases therefore provide the baseline for assessing the effect of the swimmers. At low $E$, increasing swimmer density further degrades the LCSs: the $\mathrm{FTLE}^{-}$
ridges become smeared and fragmented (Fig.~\ref{fig:ftle}(a), (d), and (g)). At intermediate $E$, however, a qualitatively different response emerges. The shrimp-free flow already exhibits substantial dislodgement of the LCSs from the underlying cellular structure (Fig.~\ref{fig:ftle}(b)). Remarkably, introducing a moderate swimmer population partially restores the original cellular organization, with the attracting ridges becoming more closely aligned with the underlying lattice (Fig.~\ref{fig:ftle}(e)). As the swimmer density is increased further, the LCSs again become more degraded; nevertheless, the dislodgement remains less pronounced than in the corresponding shrimp-free case (Fig.~\ref{fig:ftle}(h)).

At high \(E\), active matter's stabilizing effect becomes even more apparent. In the shrimp-free case, the attracting LCSs are strongly dislodged from the original cellular lattice and form highly distorted LCSs (Fig.~\ref{fig:ftle}(c)). Introducing swimmers reorganizes the large-scale LCS pattern: at moderate density, the LCSs recover a clearer cellular arrangement (Fig.~\ref{fig:ftle}(f)). Remarkably, this large-scale organization remains visible even at the highest swimmer density (Fig.~\ref{fig:ftle}(i)). Although dense swimmers introduce substantial small-scale disturbances around the ridges, the underlying LCS skeleton remains more closely aligned with the original cellular structure than in the corresponding shrimp-free flow. Thus, at sufficiently strong background forcing, increasing swimmer density does not simply degrade the LCSs; instead, the swarm can partially restore and maintain their large-scale structural organization.

To better visualize the flow topology \cite{ouellette2007curvature}, we also track the topological points of the flow (Fig.~\ref{fig:topo} and Movie~S4 of the Supplemental Material), whose panels correspond one-to-one with those of Fig.~\ref{fig:ftle}, each pair showing the same $(N,E)$ condition. The topological points are identified from local maxima of the trajectory curvature field, \(\kappa=|\mathbf{u}\times\mathbf{a}|/|\mathbf{u}|^{3}\), where \(\mathbf{u}\) and \(\mathbf{a}\) denote the tracer velocity and acceleration, respectively. Details of the extraction procedure are provided in the Supplemental Material, Sec.~S3. In the absence of swimmers, increasing the forcing causes the topological points to wander progressively farther from their reference locations on the cellular lattice, consistent with the increasing dislodgement of the LCSs (Figs.~\ref{fig:topo}(a)--\ref{fig:topo}(c)). At low forcing, increasing the swimmer density produces the same destabilizing effect: the topological points become progressively more dispersed as the swimmer density increases (Figs.~\ref{fig:topo}(a), \ref{fig:topo}(d), and \ref{fig:topo}(g)). At intermediate forcing, however, the response becomes non-monotonic. The shrimp-free flow already exhibits substantial wandering of the topological points, whereas the addition of a moderate number of swimmers causes them to condense around their reference locations, indicating stabilization of the underlying flow skeleton (Figs.~\ref{fig:topo}(b) and \ref{fig:topo}(e)). Further increasing the swimmer density reverses this trend and again broadens the distribution of the topological points, consistent with the LCS degradation observed in Fig.~\ref{fig:ftle}(h). At the highest forcing, the shrimp-free topological points are strongly dispersed, reflecting pronounced LCS dislodgement; remarkably, adding active swimmers suppresses this wandering and restores the topological points toward the cellular lattice (Figs.~\ref{fig:topo}(c), \ref{fig:topo}(f), and \ref{fig:topo}(i)).

\begin{figure*}[!t]
\centering
\includegraphics[width=1\linewidth]{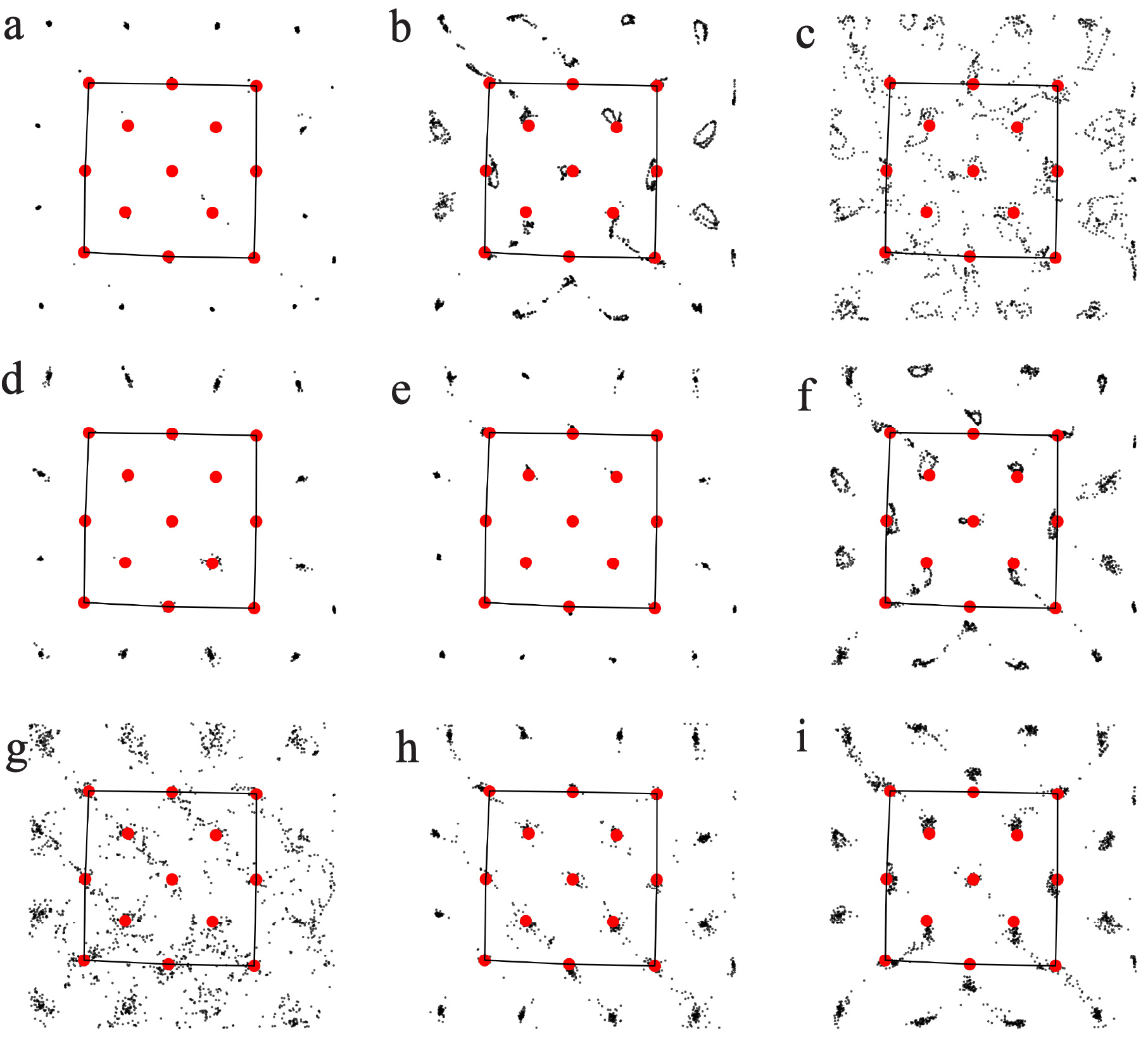}
\caption{Topological points of the flow under the same $(N,E)$ conditions as Fig.~\ref{fig:ftle}. Red dots mark the thirteen fixed points of the steady, low-Reynolds-number cellular flow imposed by the $L_m=5.1~\mathrm{cm}$ magnet lattice. The solid square simply encloses these reference points. Black dots are the instantaneous topological points extracted frame by frame from the curvature of tracer trajectories (Supplemental Material, Sec.~S3) and overlaid at fixed intervals. Tightly clustered black points sitting close to the red fixed points indicate a stable topological point array, whereas scattered black points lying far from the red references indicate a disordered, unstable array of topological points.}
\label{fig:topo}
\end{figure*}

To quantify the non-monotonic destabilizing effect from swimmers, we measure how persistent the attracting LCSs are in time. Denoting by $\mathcal{L}(t)$ the attracting-LCS region at time $t$ (the lowest $15\%$ of the $\mathrm{FTLE}^{-}$ field), the stability $R$ is the Jaccard index of two snapshots separated by one eddy turnover time,
\begin{equation}\label{eq:R}
R=\Bigl\langle \frac{|\mathcal{L}(t)\cap\mathcal{L}(t+T_e)|} {|\mathcal{L}(t)\cup\mathcal{L}(t+T_e)|} \Bigr\rangle_t,
\end{equation}
where $|\cdot|$ denotes area and $\langle\cdot\rangle_t$ averages over all snapshot pairs in a run. By construction, $0\le R\le1$: $R = 1$ corresponds to an LCS that is nearly unchanged over a turnover time, and $R = 0$ to one that reorganizes completely. The construction is illustrated in Figs.~\ref{fig:stability}(b--d), where the example shown has $R=0.12$, and in Movie~S3 of the Supplemental Material. We use the Jaccard index rather than a direct correlation of the full $\mathrm{FTLE}^{-}$ fields because the latter conflates skeleton stability with the small-scale fluctuations the swimmers inject; the Supplemental Material (Sec.~S2.3) shows that neither $CC$ nor $R$ depends sensitively on the choice of threshold.

Figure~\ref{fig:stability}(a) maps $R$ across the $(E,N)$ plane, with the triangles marking the nine representative conditions of Figs.~\ref{fig:ftle} and~\ref{fig:topo}.
In the shrimp-free flow ($N=0$) $R$ stays between $0.55$ and $0.64$ at the three lowest forcings ($E\le0.13~\mathrm{cm^2\,s^{-2}}$), where the cellular array is steady. It drops to $0.2$ between $E=0.13$ and $0.30~\mathrm{cm^2\,s^{-2}}$, where the array becomes time-dependent, and decreases slowly beyond as the pattern evolves toward weak turbulence.
At moderate swimmer density ($N\approx50$--$150$), however, $R$ first rises and then falls as $E$ increases, so that the LCSs are most persistent at an intermediate forcing rather than at the lowest one. The swimmers stabilize the LCSs over a band of $E$, and the forcing at which stability peaks is shifted to higher $E$ relative to the dilute case. At a higher density, this peak continues to shift toward higher $E$ while the overall stability decreases, as the increasingly dense swarm disorders the LCSs across the whole range. Reading the map at fixed forcing gives a consistent picture: over an intermediate band of $E$, $R$ is non-monotonic in $N$, rising to a maximum at finite density before falling again. The binned curves of Fig.~\ref{fig:stability}(e--n) confirm this statistically: the curves change from decreasing with $E$ at $N=0$ to peaked at finite density.

Because the stabilization is the more surprising of the two responses, we checked that it is not a measurement artifact. The most plausible candidate is the masking of the swimmers before the PTV step (Sec.~\ref{sec:ptv}), which leaves more holes in the interpolated velocity field. Such an artifact would grow monotonically with $N$. A density-dependent smoothing cannot produce a stabilization that appears only above a threshold in a variable that does not enter the masking. As an additional control, we artificially increased the masked area in the low-density cases to mimic the amount of masking present at higher swimmer densities. Removing these additional spatial regions produced no appreciable change in the reconstructed LCSs (Supplemental Material, Sec.~S2.2). This further indicates that the observed stabilization is not a consequence of the increased masking associated with denser swimmer populations.

\begin{figure*}[!t]
\centering
\includegraphics[width=0.9\linewidth]{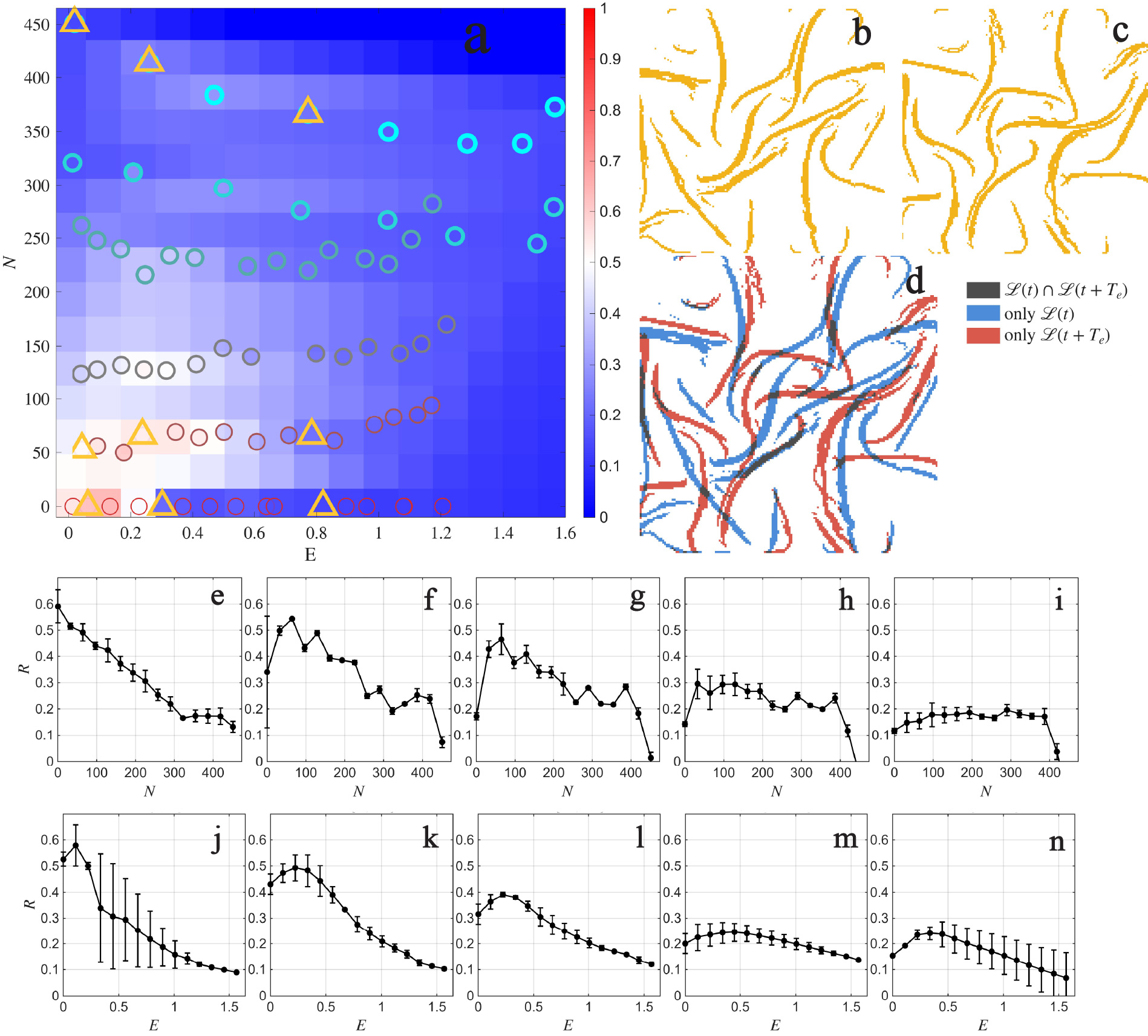}
\caption{Temporal stability $R$ of the flow skeleton (Eq.~\ref{eq:R}) across the $(E,N)$ parameter space, where $E$ is the measured flow energy of each condition (Eq.~\ref{eq:E}) and $N$ the swimmer count. \textbf{(a)} Heat map of $R$; the background color shows $R$, obtained by interpolation. Circles mark all measured conditions and triangles mark the nine representative cases of Figs.~\ref{fig:ftle} and \ref{fig:topo}; the symbols are colored separately by swimmer count $N$, from red ($N=0$) to cyan (highest $N$).  \textbf{(b--d)} Illustration of the stability metric for a single condition. \textbf{(b)} The attracting-LCS region $\mathcal{L}(t)$ at one instant and \textbf{(c)} the region $\mathcal{L}(t+T_e)$ one eddy turnover time later (yellow, the lowest $15\%$ of the $\mathrm{FTLE}^{-}$ field). \textbf{(d)} Their overlap: the intersection $\mathcal{L}(t)\cap\mathcal{L}(t+T_e)$ (grey), the part only in $\mathcal{L}(t)$ (blue), and the part only in $\mathcal{L}(t+T_e)$ (red). \textbf{(e--i)} $R$ versus $N$ within fixed $E$ bins: $E\in[0,0.2]$, $[0.2,0.4]$, $[0.4,0.6]$, $[0.6,1.0]$, and $[1.0,1.4]$. \textbf{(j--n)} $R$ versus $E$ within fixed $N$ bins: $N\in[0,35]$, $[35,150]$, $[150,250]$, $[250,350]$, and $[350,450]$. The binned curves in (e--n) are evaluated on the interpolated field of (a), markers and error bars give the mean and standard deviation of the interpolant within each bin.}
\label{fig:stability}
\end{figure*}

A possible explanation for the non-monotonic destabilization behavior is that, when the swarm is weakly organized by the background flow, the swimmers remain approximately randomly distributed and oriented, so their individual jet-like wakes perturb the flow in largely random directions throughout the domain. Such incoherent agitation tends to displace the strongly attracting material lines and to drive the topological points away from the lattice imposed by the electromagnetic forcing. This behavior is observed at the lowest forcing, where \(K_s/E\) is largest: \(R\) decreases monotonically with \(N\) (Fig.~\ref{fig:stability}(e)), while the topological points become increasingly scattered (Fig.~\ref{fig:topo}(g)).

As the background flow becomes stronger, however, it can increasingly reorient and spatially organize the swimmers. The swimmers then preferentially accumulate near the attracting LCSs and align with their local direction. Their jet-like wakes will therefore become correlated with the underlying coherent structure rather than acting as random perturbations, reinforcing the LCSs and producing the observed stabilization.
This correspondence suggests that flow-induced organization of the swimmers may be linked to the stabilization of the LCSs. Establishing the mechanism, however—particularly why stabilization is strongest at intermediate swimmer density and intermediate forcing—requires direct measurements of swimmer orientation and wake–strain coupling, and is left for future study.

\section{Summary and conclusions}\label{sec:conclusion}

We have experimentally investigated the two-way coupling between a swarm of centimeter-scale swimmers and a structured background flow. Brine shrimp (\textit{A. salina}) were introduced into an electromagnetically driven quasi-2D cellular flow, and the swimmer density and the background forcing were varied independently across the full $(N,E)$ parameter space. Using a set of Lagrangian diagnostics built on the attracting LCSs and on the topological points of the flow, we characterized both how the swimmers are organized within the flow and how they reshape the flow in return.

We first quantified the accumulation of swimmers onto the attracting LCSs. In the dilute limit, our measurements reproduce the mobility-selective accumulation predicted by the one-way-coupled simulations \citep{si2021}: the accumulation onto the attracting LCSs is strongest when the swimming speed is comparable to the flow speed. As the swarm becomes denser, however, the mobility selectivity is progressively lost.

We then examined the back-action of the swimmers on the LCSs. A dense swarm in a weakly forced flow degrades the LCSs, as reported previously \citep{si2024}. When the forcing is strong enough to organize the swarm, the effect reverses: the LCSs are more persistent at finite density than in the shrimp-free flow, and retain the cellular organization imposed by the magnet array even where the unladen flow has already lost it. The forcing at which LCSs start to dislodge shifts to higher values as the swarm grows. The FTLE-based and topological diagnostics agree. To our knowledge, such stabilization by active matter composed of centimeter-scale swimmers has not been reported previously.

These results show that the two directions of the coupling are not independent: the swimmers are organized by LCSs, and they modify those structures in turn. Both directions have previously been studied only in the one-way limit, which we find incomplete in either direction.

Several questions remain open. First, our understanding of the differences between one-way and two-way coupling remains partly inferential. In particular, further investigation is needed to clarify how the background flow organizes the spatial distribution of swimmers and how the swimmers, in turn, reshape the LCSs that form the flow skeleton. Second, extending the present study to stronger, fully turbulent flows and ultimately to three-dimensional systems represents an important direction for future research on two-way coupling.

\bibliography{PRF}

\end{document}


\begin{center}
{\Large \textbf{Supplemental Material for}}\\[6pt]
{\large \textbf{Two-way coupling in active suspensions suppresses mobility-selective accumulation and induces non-monotonic flow stabilization}}\\[12pt]

Ziyue Yu$^{\ast}$,
Aineo M. Wu,
Lei~Fang$^{\ast}$\\
\small$^\ast$Corresponding authors. Email: lei.fang@pitt.edu; ZIY79@pitt.edu
\end{center}

\subsection*{This PDF file includes:}
\begin{itemize}
    \item Experimental method and analysis details (Secs.~S1 to S4)
    \item Figures S1 to S6
    \item Table S1
    \item Captions for Movies S1 to S4
\end{itemize}

\newpage

\section{Experimental method}
\label{sisec:method}

\subsection{Apparatus}
\label{sisec:apparatus}
The quasi-2D flow was generated in the electromagnetically driven thin-layer
system used in our previous work
\citep{kelley2011,fang2018boundaries,si2022,si2024,yu2026tensor}. An acrylic frame (96.5 cm × 83.8 cm) holds a tempered glass plate (81.3 cm × 81.3 cm), with a pair of copper electrodes mounted on opposite sides. The electrolyte layer rests on the glass, and the magnet array lies beneath it. The top surface of the glass was coated with a hydrophobic agent (Rain-X) to reduce friction, and the bottom surface was covered with a light-absorbing film to suppress reflections.

\begin{figure}[H]
\centering
\includegraphics[width=0.45\linewidth]{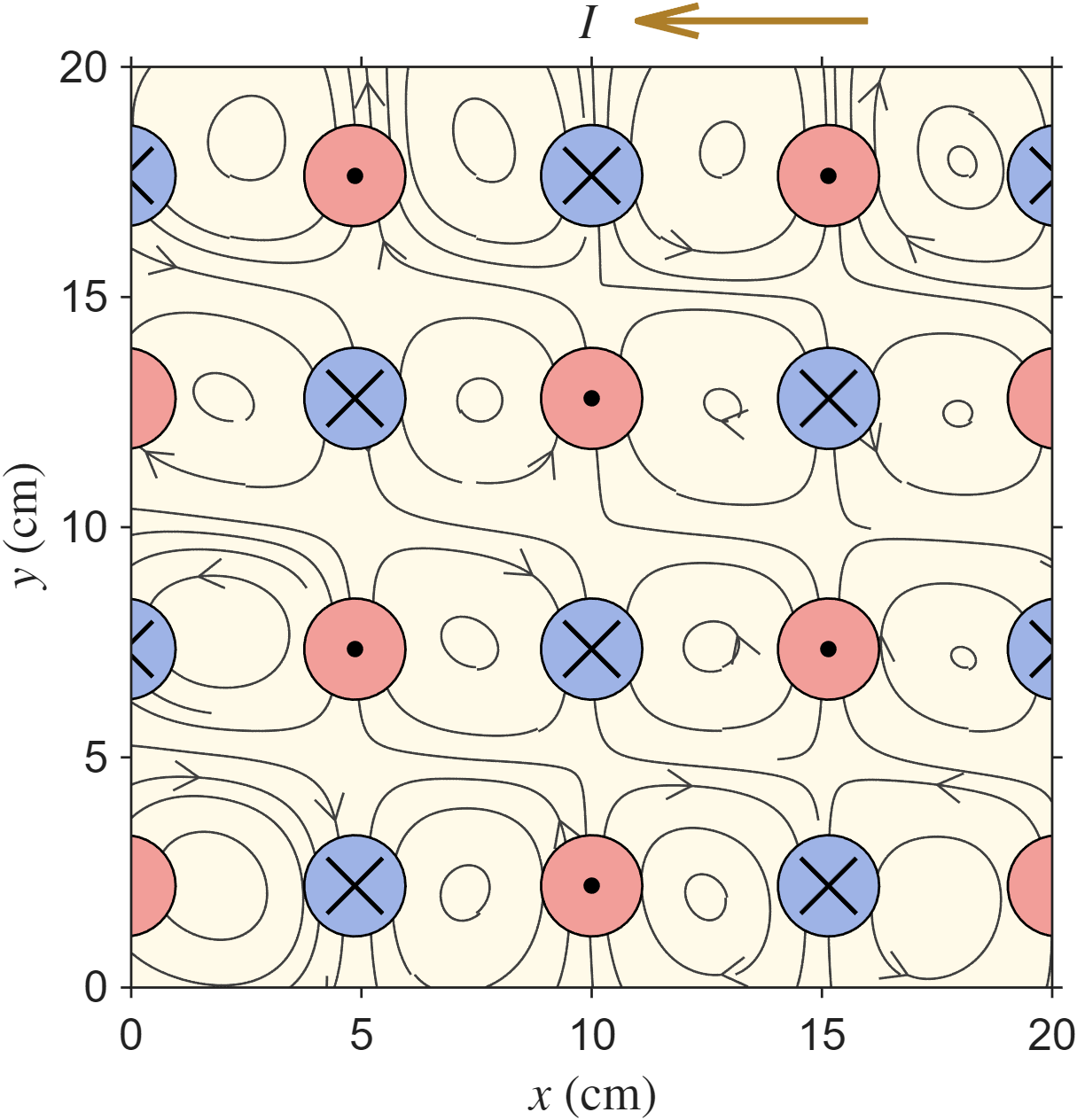}
\caption{Top view of the $20\times20~\mathrm{cm}^2$ imaging window with
the magnet lattice beneath it and the flow it drives. The curves are
streamlines of the time-averaged velocity field of the shrimp-free run at
the lowest forcing, the field of main Fig.~1(b). The lattice is drawn at its spacing $L_m=5.1~\mathrm{cm}$ and centered in the window; the magnet symbols are enlarged for clarity. A dot marks a magnet whose jet is along $+y$ and a cross one
whose jet is along $-y$.}
\label{fig:setuptop}
\end{figure}

Cylindrical neodymium magnets (grade N52; diameter $1.27~\mathrm{cm}$,
thickness $0.64~\mathrm{cm}$) were placed beneath the glass in a checkerboard
arrangement with alternating polarity and a center-to-center spacing of
$L_m=5.1~\mathrm{cm}$. Figure~\ref{fig:setuptop} shows the lattice under
the imaging window together with the flow it drives.

A $6~\mathrm{mm}$ layer of $14\%$ NaCl solution by mass (density
$\rho=1.101~\mathrm{g\,cm^{-3}}$, kinematic viscosity
$\nu=1.25\times10^{-2}~\mathrm{cm^2\,s^{-1}}$) was deposited on the glass. A
direct current between the electrodes interacts with the magnetic field
to produce a Lorentz force that alternates in sign from magnet to magnet and drives
the square array of counter-rotating vortices. The forcing strength was controlled by the current, which ranged from 0.1 to 2.3 A. Since the depth of the fluid layer $h$ is small compared with the injection length scale $L_m$ ($h/L_m \approx 0.12$), the flow remained effectively two-dimensional over this range \citep{kelley2011}.

\subsection{Active matter: \textit{Artemia salina}}
\label{sisec:animals}
The swimmers are adult brine shrimp, \textit{Artemia salina}, obtained from a
commercial supplier. Body length and width were
measured directly with a ruler: the body length was approximately
$1~\mathrm{cm}$ and the aspect ratio (length to width) was
about $3$, with the width taken across the appendages. The shrimp were held in $3\%$ NaCl solution for
$24~\mathrm{h}$ before use. Since they are less dense than the working fluid, they floated
at the free surface of the layer.

\begin{figure}[H]
\centering
\includegraphics[width=0.85\linewidth]{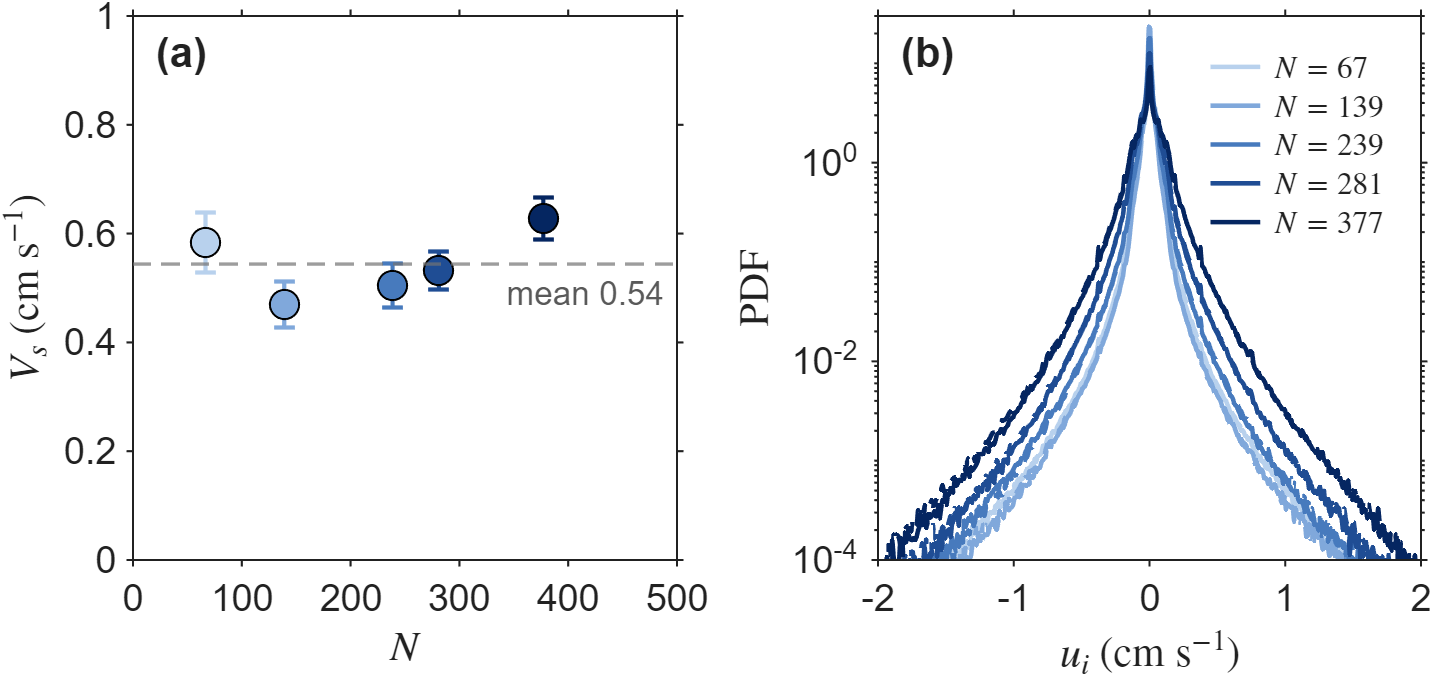}
\caption{Swimming speed of \textit{A.~salina} and the flow driven by the
swarm in the absence of background forcing. \textbf{(a)} Mean swimming speed
in the quiescent run of each loading; error bars give the standard
deviation over time of the per-frame mean speed, and the dashed line marks
the mean $V_s$ over the five runs. \textbf{(b)} Probability density
functions of the velocity components $u$ (solid lines) and $v$ (dashed
lines) of the swarm-induced flow in the quiescent run of each loading, from
the PTV track velocities; the color darkens with $N$
(Table~\ref{tab:conditions}).}
\label{fig:shrimp}
\end{figure}

The shrimp swimming speed was measured in the quiescent runs by tracking the centroid of each shrimp
from frame to frame (Sec.~\ref{sisec:shrimp}): the centroids of consecutive
frames are paired by nearest-neighbor matching, and the
speed is the paired displacement divided by the frame interval.
Figure~\ref{fig:shrimp}(a) shows the mean speed in the quiescent run of
each loading. The
mean over the runs is $V_s=0.54~\mathrm{cm\,s^{-1}}$ with a run-to-run
standard deviation of $0.06~\mathrm{cm\,s^{-1}}$ and no systematic dependence
on $N$, so $V_s$ is treated as a constant in the main text. Under the uniform
illumination used in the experiments, the shrimp show no directional bias.
Figure~\ref{fig:shrimp}(b) shows the probability density functions (PDFs)
of the two velocity components of the flow driven by the swarm alone in the
quiescent runs, built from the tracer velocities of the PTV tracks
(Sec.~\ref{sisec:ptv}). At each loading the PDFs of $u$ and $v$ are nearly
identical, indicating that the swarm-induced flow is statistically isotropic.

\subsection{Detection, masking, and counting of the shrimp}
\label{sisec:shrimp}

Before particle tracking velocimetry (PTV), the shrimp were separated from the tracers by image processing. This separation relies on their large size difference: the shrimp body length is about two orders of magnitude larger than the tracer diameter.

First, the images were binarized, and then the regions larger than an area threshold set by the body size of the shrimp were identified as shrimp. Second, all pixels of every such region were set to the background intensity, so that the shrimp were excluded from the subsequent PTV.

The number of shrimp $N$ in a run is the number of removed regions, averaged over the frames of the run, and the centroid of each region
gives the position of the shrimp used for the accumulation coefficient $CC$
and for the swimming-speed measurement of Sec.~\ref{sisec:animals}. Removing
the shrimp leaves gaps in the tracer particle data, which are filled when the particle velocities are interpolated onto the Eulerian grid; this does not affect the LCSs extracted
from the field, as shown by the masking control of
Sec.~\ref{sisec:maskrobust}.

\subsection{Particle tracking velocimetry}
\label{sisec:ptv}
The flow was seeded with green fluorescent polyethylene tracer particles
(Cospheric; density $1.0125~\mathrm{g\,cm^{-3}}$, diameter
$106$--$125~\mu\mathrm{m}$). Being slightly less dense than the working fluid,
the tracers remain at the free surface and follow the surface flow. A small
amount of surfactant was added to prevent surface-tension-driven clustering. The tracer Stokes number is
\begin{equation}\label{eq:St}
  St=\frac{1}{18}\frac{\rho_p}{\rho_f}\frac{d^2}{L_m^2}Re ,
\end{equation}
which lies between $1\times10^{-5}$ and $2\times10^{-4}$ over the range
$Re=40$--$700$ of the experiments, so the tracers follow the fluid motion
faithfully \citep{ouellette2008transport}.

The central $20\times20~\mathrm{cm}^2$ region of the layer was imaged from
above with a machine vision camera (Basler \texttt{acA2040-90um}) at
$1600\times1600$ pixels and $60$ frames per second, after the background flow
(and the swarm, when present) had reached a statistically stationary state.
Between $2\times10^{4}$ and $6\times10^{4}$ tracers were tracked per frame. Tracer positions and velocities were obtained with the PTV algorithm of
Ouellette \emph{et al.}~\citep{ouellette2006ptv}. The tracer velocities were interpolated linearly onto a regular Eulerian grid. All spatial averages and statistics were computed over this full
$20\times20~\mathrm{cm}^2$ domain.

\subsection{Experimental conditions}
\label{sisec:protocol}
The shrimp-free background flow was recorded first, at each dc current in
turn.
Shrimp were then added in increments. For each loading, the flow driven by the swarm alone was first recorded with the current switched off (the quiescent run), and the current sweep was then repeated with the swarm present.
Every record was started after the flow
and the swarm had reached a statistically stationary state. Table~\ref{tab:conditions} summarizes
the conditions. The number of shrimp $N$ is the time-averaged count of
Sec.~\ref{sisec:shrimp}; it varies from condition to condition within a
loading because shrimp occasionally leave or enter the field of view, and
the spread quoted in the table is the standard deviation over the conditions
of that loading.
The kinetic energy of the flow driven by the swarm alone,
$K_s=\tfrac12\langle\mathbf{u}_s\cdot\mathbf{u}_s\rangle$ evaluated in a separate series of calibration runs with the current switched off, grows linearly with $N$ (main Fig.~1(d)). Here $\mathbf{u}_s$ denotes the velocity field induced by the swarm.

\begin{table}[H]
\centering
\caption{Experimental conditions. $N$ is the mean $\pm$ standard deviation of
the number of shrimp over the conditions of each loading; $I$ is the nominal
dc current and $\Delta I$ its increment; $E$ is the mean kinetic energy per
unit mass measured in each condition. The quiescent run
of each loading, used to calibrate $V_s$, is not counted among the
conditions.}
\label{tab:conditions}
\begin{tabular}{lccccc}
\toprule
Loading & $N$ & $I$ (A) & $\Delta I$ (A) & Conditions & $E$ ($\mathrm{cm^2\,s^{-2}}$) \\
\midrule
Shrimp-free & $0$         & $0.1$--$1.6$ & $0.1$ & 16 & $0.014$--$1.21$ \\
1           & $67\pm13$   & $0.1$--$1.6$ & $0.1$ & 16 & $0.010$--$1.17$ \\
2           & $139\pm12$  & $0.1$--$1.6$ & $0.1$ & 16 & $0.008$--$1.22$ \\
3           & $239\pm17$  & $0.1$--$1.6$ & $0.1$ & 16 & $0.010$--$1.17$ \\
4           & $281\pm27$  & $0.2$--$2.3$ & $0.3$ & 8  & $0.012$--$1.56$ \\
5           & $377\pm39$  & $0.2$--$2.3$ & $0.3$ & 8  & $0.019$--$1.57$ \\
\bottomrule
\end{tabular}
\end{table}

\section{Lagrangian coherent structures and sensitivity analysis}
\label{sisec:lcs}
\subsection{Flow-map integration and the FTLE field}
\label{sisec:ftle}

The LCSs are detected from the finite-time Lyapunov exponent (FTLE),
following \citet{shadden2005} and \citet{haller2015}. All Lagrangian fields
are built from the flow map
\begin{equation}\label{sieq:flowmap}
\mathbf{F}_{t_0}^{t}(\mathbf{x}_0)=\mathbf{x}(t;t_0,\mathbf{x}_0), \qquad
\mathbf{x}(t;t_0,\mathbf{x}_0)=\mathbf{x}_0+\int_{t_0}^{t}\mathbf{u}(\tau,\mathbf{x}(\tau;t_0,\mathbf{x}_0))\,\mathrm{d}\tau ,
\end{equation}
which carries a fluid element from its position $\mathbf{x}_0$ at time $t_0$
to its position $\mathbf{x}$ at time $t$. Here $\mathbf{u}$ is the gridded
velocity field of Sec.~\ref{sisec:ptv} and the initial positions
$\mathbf{x}_0$ are the points of the Eulerian grid. The integral is evaluated
with the explicit second-order Runge--Kutta scheme. For the attracting LCSs the integration runs backward in time
($t<t_0$).

The gradient of the flow map is approximated by central differences over the
neighboring grid points,
\begin{equation}\label{sieq:gradF}
\nabla\mathbf{F}_{t_0}^{t}(\mathbf{x}_0)\approx
\begin{bmatrix}
\dfrac{x(t;t_0,\mathbf{x}_0+\boldsymbol{\delta}_1)-x(t;t_0,\mathbf{x}_0-\boldsymbol{\delta}_1)}{2|\boldsymbol{\delta}_1|} &
\dfrac{x(t;t_0,\mathbf{x}_0+\boldsymbol{\delta}_2)-x(t;t_0,\mathbf{x}_0-\boldsymbol{\delta}_2)}{2|\boldsymbol{\delta}_2|}\\[1.5ex]
\dfrac{y(t;t_0,\mathbf{x}_0+\boldsymbol{\delta}_1)-y(t;t_0,\mathbf{x}_0-\boldsymbol{\delta}_1)}{2|\boldsymbol{\delta}_1|} &
\dfrac{y(t;t_0,\mathbf{x}_0+\boldsymbol{\delta}_2)-y(t;t_0,\mathbf{x}_0-\boldsymbol{\delta}_2)}{2|\boldsymbol{\delta}_2|}
\end{bmatrix},
\end{equation}
where $\boldsymbol{\delta}_1$ and $\boldsymbol{\delta}_2$ are the grid
vectors along $x$ and $y$. The right Cauchy--Green strain tensor
\begin{equation}\label{sieq:CG}
\mathbf{C}(\mathbf{x}_0)=\big(\nabla\mathbf{F}_{t_0}^{t}\big)^{\mathsf{T}}\nabla\mathbf{F}_{t_0}^{t}
\end{equation}
describes how an initially small perturbation of $\mathbf{x}_0$ has grown by
time $t$. It is symmetric and positive definite, with eigenvalues
$0<\lambda_1\le1\le\lambda_2$, and the FTLE is
\begin{equation}\label{sieq:ftle}
\mathrm{FTLE}(\mathbf{x}_0)=\frac{1}{t-t_0}\,\tfrac{1}{2}\ln\lambda_2(\mathbf{x}_0).
\end{equation}
For backward integration $t-t_0<0$, so the backward field $\mathrm{FTLE}^{-}$
is negative, and its ridges, the regions of locally most negative values,
mark the attracting LCSs \citep{haller2015}.

The integration time is one eddy turnover time $T_e=L_m/U$, evaluated for
each condition from its own measured $U$, so that every condition is
compared after the same amount of its own dynamical time.

\subsection{Effect of the shrimp masks on the LCSs}
\label{sisec:maskrobust}
The shrimp are masked to the background intensity in the particle images before tracking
(Sec.~\ref{sisec:shrimp}), so the velocity in the masked regions is
supplied by the interpolation of Sec.~\ref{sisec:ptv}, and the masked
fraction of the image grows with $N$, from $0.8\%$ at $N\approx70$ to
$5.0\%$ at $N\approx410$. The LCSs, and every quantity built on them,
might therefore depend on $N$ through the masked area alone. This section
shows that the masks have a negligible effect on the LCSs, and hence on $R$.

\begin{figure}[H]
\centering
\includegraphics[width=\linewidth]{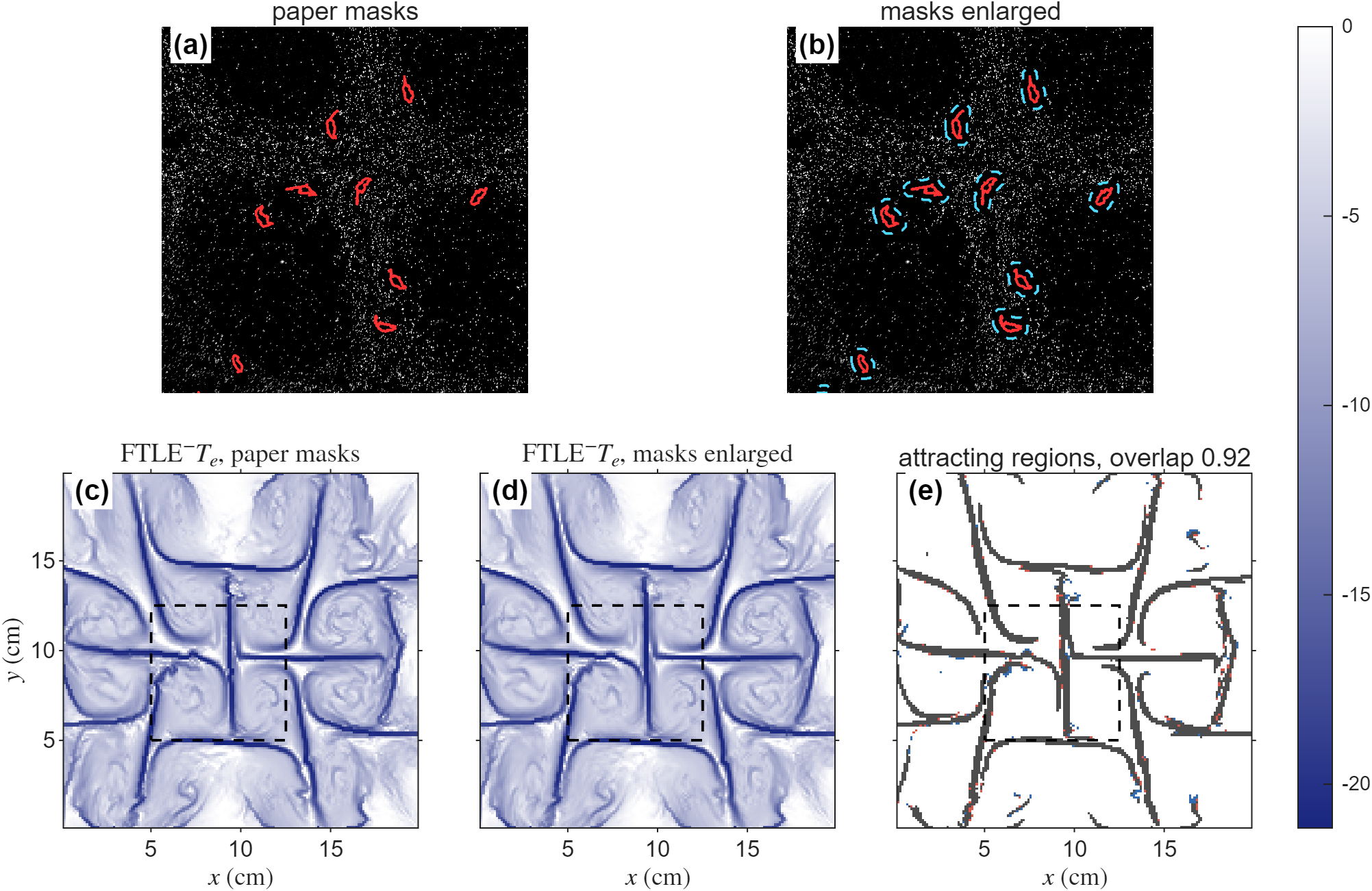}
\caption{Masking control. \textbf{(a,b)} A $7.5\times7.5~\mathrm{cm}^2$
part of one particle image of the run at $N\approx70$ and
$E=0.34~\mathrm{cm^2\,s^{-2}}$ (loading~1 in Table~\ref{tab:conditions}) with the masks used in the main text (red)
and the same masks dilated to six times their area (blue dashed).
\textbf{(c,d)} $\mathrm{FTLE}^{-}T_e$ of the same snapshot over the full
field of view, computed from (a) and from (b). \textbf{(e)} Attracting-LCS
regions of (c) and (d): grey where the two coincide, blue only in (c), red
only in (d). The dashed square in (c--e) marks the part of the field shown
in (a,b), which is displayed in the same orientation.}
\label{fig:maskctrl}
\end{figure}

To test whether the masked regions influence the LCSs and the quantities
derived from them, two cases were compared on one run of loading~1
($N\approx70$, $E=0.34~\mathrm{cm^2\,s^{-2}}$):
(i) the shrimp masks as used in the main text, which cover $0.8\%$ of the
image (Fig.~\ref{fig:maskctrl}(a));
and (ii) the same masks dilated so that their total area increases by a
factor of about six and the masked fraction rises to $4.8\%$
(Fig.~\ref{fig:maskctrl}(b)).
In both cases the masked particle images were processed with the identical
tracking, gridding, and FTLE chain of Secs.~\ref{sisec:ptv} and
\ref{sisec:ftle}, and the LCSs of case (ii) were compared with those of
case (i).

The masks do not change the LCSs. The $\mathrm{FTLE}^{-}$ field of case (ii) has a correlation of $0.986$ with that of case (i),
and the attracting-LCS regions of the two overlap with a Jaccard index of
$0.92\pm0.01$ (Figs.~\ref{fig:maskctrl}(c--e)). The reason is that
the masked area is small, at most $5\%$ of the image even for the densest
swarm, and the velocity inside each mask is interpolated from the tracers
around it. The flow map integrates this velocity along each trajectory over
one $T_e$, during which a tracer travels a distance of order $L_m$, far
larger than a mask, so the local interpolation inside the masks is averaged
out and the LCSs, which reflect the overall motion of the tracers, remain
the same.

Because $R$ is the Jaccard overlap between the attracting-LCS regions of two
snapshots one $T_e$ apart, an unchanged LCS region leaves $R$ unchanged.
Case (i) gives $R=0.568$; case (ii) gives $R=0.595$.
The effect of masking on $R$ is therefore negligible. The densest loading at the same forcing lowers $R$ from $0.56$ to $0.27$, nearly ten times the masking effect, so the dependence of $R$ on $N$ reported in the main
text is not produced by the masked area.

\subsection{Sensitivity of $CC$ and $R$ to the LCS threshold}
\label{sisec:threshold}
Both the accumulation coefficient $CC$ and the stability $R$ of the main
text are built on the attracting-LCS region, defined as the lowest $15\%$ of
each $\mathrm{FTLE}^{-}$ snapshot. To check that the conclusions do not
depend on this choice, both quantities were recomputed with the region
defined by the lowest $10\%$, $15\%$, and $20\%$ of the field, for every
condition of Table~\ref{tab:conditions}.

\begin{figure}[H]
\centering
\includegraphics[width=\linewidth]{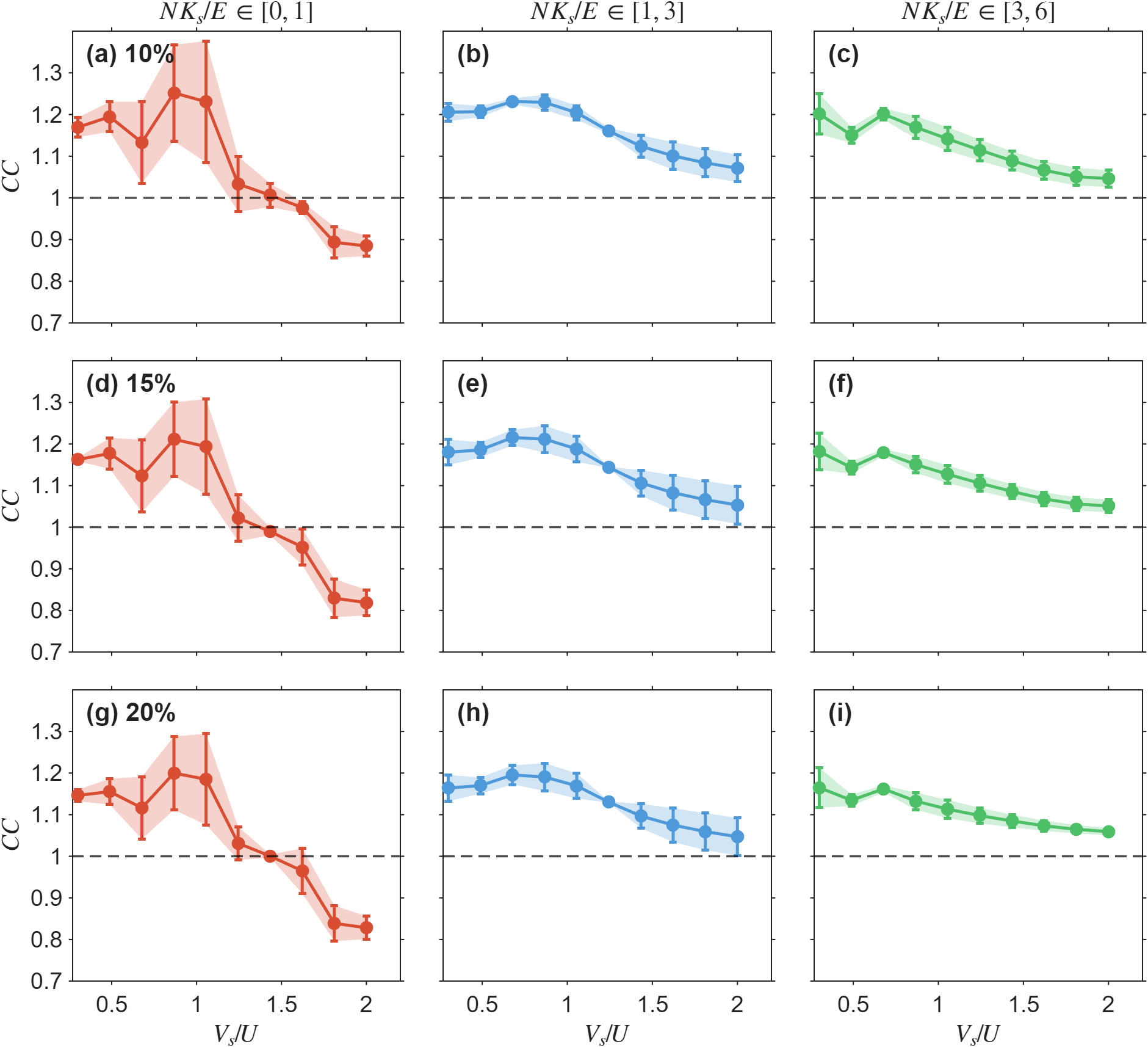}
\caption{ Threshold sensitivity of the accumulation coefficient. $CC$ versus
$V_s/U$ in the three $NK_s/E$ bins of main Fig.~2(c--e), with the
attracting-LCS region defined by the lowest \textbf{(a--c)} $10\%$,
\textbf{(d--f)} $15\%$, and \textbf{(g--i)} $20\%$ of the
$\mathrm{FTLE}^{-}$ field. Markers and error bars give the mean and standard
deviation of the interpolated field within each bin, as in main
Fig.~2(c--e), to which (d--f) correspond; the dashed line marks $CC=1$.}
\label{fig:ccthr}
\end{figure}

Figure~\ref{fig:ccthr} shows $CC$ versus $V_s/U$ in the three $NK_s/E$
ranges of main Fig.~2(c--e), with the attracting-LCS region defined by the
lowest $10\%$, $15\%$, and $20\%$ of the $\mathrm{FTLE}^{-}$ field. The
three thresholds give the same result as the main text. At the lowest
$NK_s/E$, $CC$ peaks
near $V_s/U\approx1$: the accumulation onto the attracting LCSs is
strongest when the swimming speed is comparable to the flow speed. At larger
$NK_s/E$ the peak flattens and $CC$ becomes nearly independent of $V_s/U$,
so the mobility selectivity is lost while no depletion ($CC<1$) is measured outside the most dilute loading.

\begin{figure}[H]
\centering
\includegraphics[width=\linewidth]{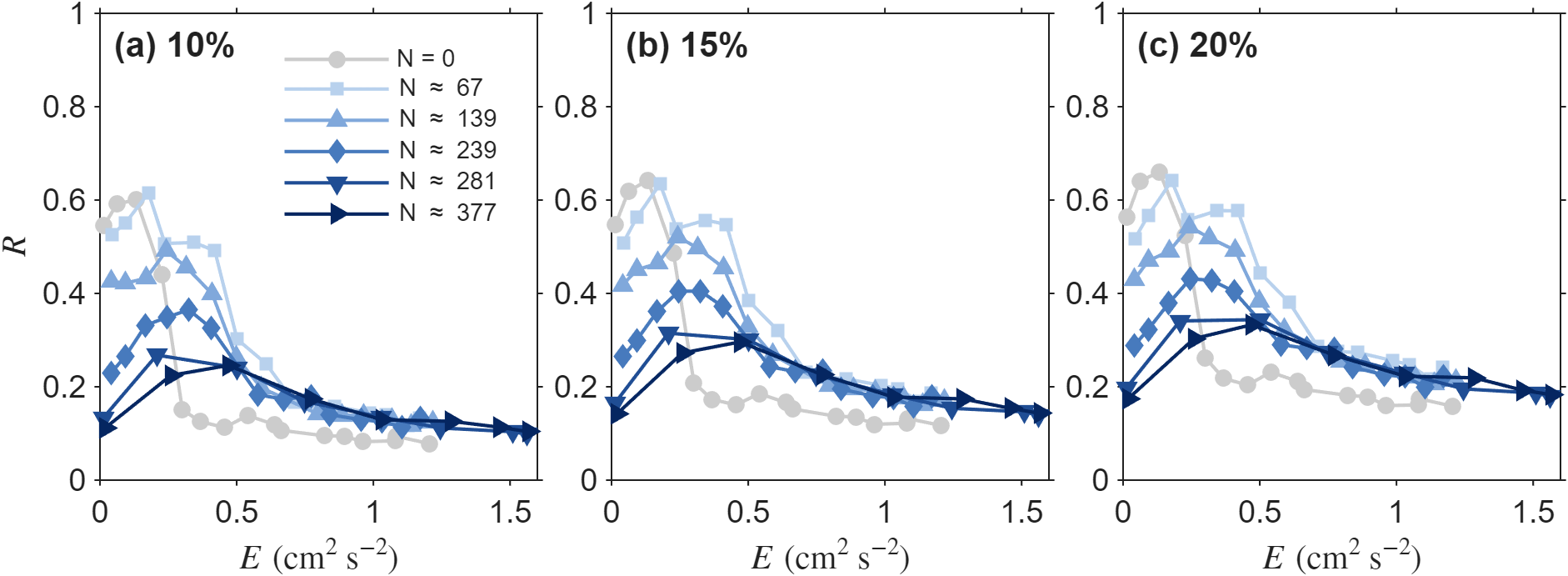}
\caption{ Threshold sensitivity of the stability $R$. \textbf{(a--c)} $R$
versus the background forcing $E$ for the shrimp-free flow and the five
loadings of Table~\ref{tab:conditions}, with the attracting-LCS region
defined by the lowest $10\%$, $15\%$, and $20\%$ of the $\mathrm{FTLE}^{-}$
field. Panel (b) is the definition used in the main text.}
\label{fig:rthr}
\end{figure}

Figure~\ref{fig:rthr} shows the result for $R$. The dependence on $E$ and $N$ is unchanged: at
every threshold $R$ in the shrimp-free flow is highest at the three lowest forcings and falls steeply between $E=0.13$ and $0.30~\mathrm{cm^2\,s^{-2}}$, is peaked at intermediate $E$ for the swarm, and exceeds the shrimp-free value over the same band of $E$.

The heat maps of main Figs.~2(a) and 5(a) are natural-neighbor
interpolations of the values measured at the individual conditions, which
are marked by the symbols; the interpolant is a smooth function of the
measured points and does not create features absent from them.

\section{Extraction of the topological points}
\label{sisec:topo}

\begin{figure}[H]
\centering
\includegraphics[width=\linewidth]{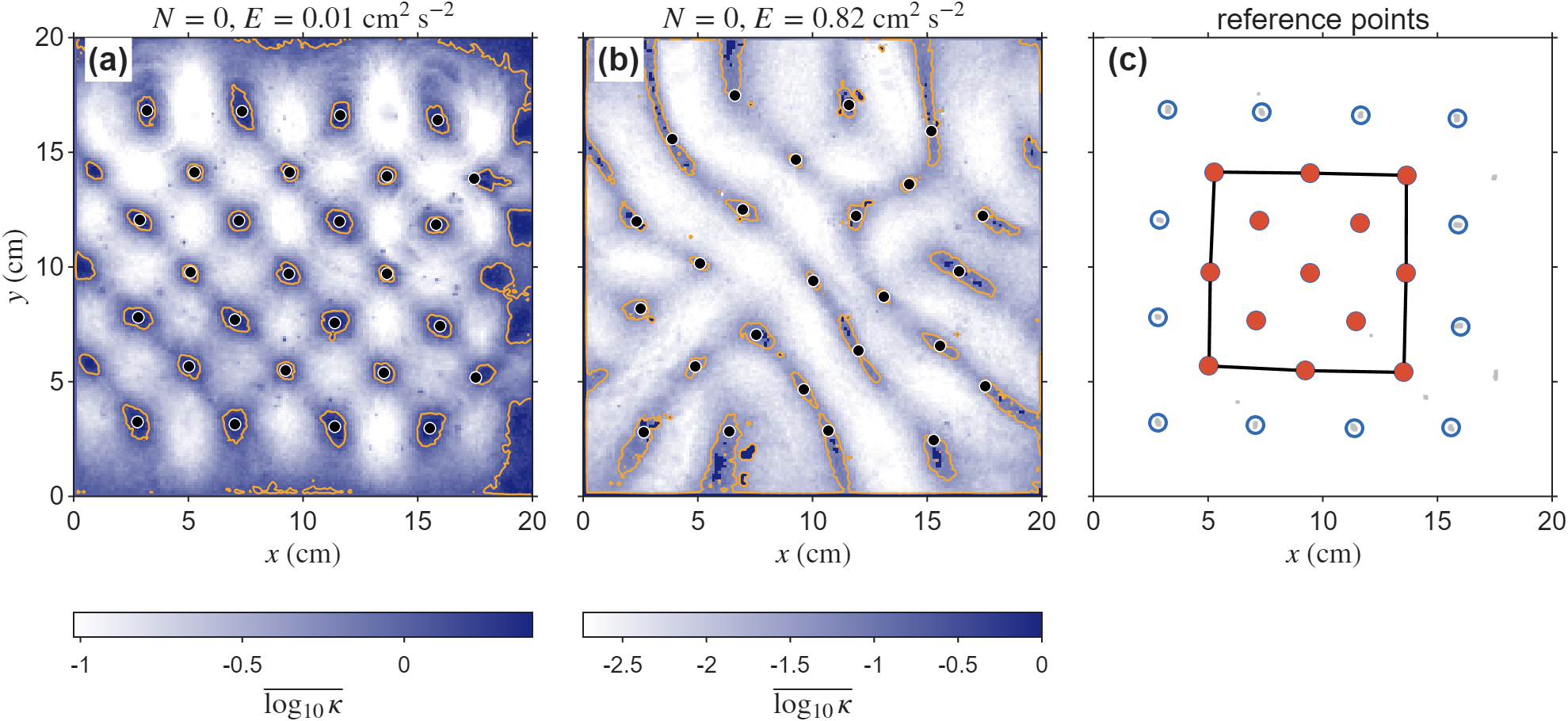}
\caption{Extraction of the topological points. \textbf{(a)} The
time-smoothed curvature field $\overline{\log_{10}\kappa}$ of one frame of
the shrimp-free run at the lowest forcing. The orange contour bounds the
regions of highest curvature and the black points are the curvature-weighted
centroids of the regions inside the contour. \textbf{(b)} The same for the shrimp-free
run at $E=0.82~\mathrm{cm^2\,s^{-2}}$, the condition of main Fig.~4(c).
\textbf{(c)} All points extracted from the
run of (a) (grey), the centers of their $25$ clusters (blue circles), and the
thirteen retained as the reference points of main Fig.~4 (red) with their
convex hull.}
\label{fig:topoext}
\end{figure}

The topological points of main Fig.~4 are located from the curvature of the
tracer trajectories \citep{ouellette2007curvature}. Near a vortex center or
a saddle the tracers turn most sharply, so the trajectory curvature
\begin{equation}\label{eq:curvature}
  \kappa=\frac{|\mathbf{u}\times\mathbf{a}|}{|\mathbf{u}|^{3}}
\end{equation}
is largest there; $\mathbf{u}$ and $\mathbf{a}$ are the velocity and
acceleration of a tracer. In every frame, $\kappa$ is evaluated for each
tracked particle from the PTV velocity and acceleration of
Sec.~\ref{sisec:ptv}, and the particle values are interpolated onto the
Eulerian grid of that section. Because $\kappa$ spans several orders of
magnitude and is sensitive to a few extreme values, the field is processed
in logarithmic form: the outliers of $\log_{10}\kappa$ in each frame are
discarded, and the field is smoothed over a time window short compared with
$T_e$, giving $\overline{\log_{10}\kappa}$ (Fig.~\ref{fig:topoext}(a,b)).
The connected regions in which $\overline{\log_{10}\kappa}$ is highest are
the topological points, each reduced to one point, its curvature-weighted
centroid.

The points are not classified into centers and saddles: main Fig.~4 uses
only their number and their positions. In the steady cellular flow
(Fig.~\ref{fig:topoext}(a)) the procedure returns one point per vortex
center and per saddle, on the lattice of the magnet array. At strong
forcing (Fig.~\ref{fig:topoext}(b)) the regions of high curvature are
displaced and distorted, and the points leave the lattice.

The reference points of main Fig.~4 (red) come from the shrimp-free run at
the lowest forcing ($E=0.01~\mathrm{cm^2\,s^{-2}}$), in which the flow is
steady. The points extracted from all frames of that run fall into $25$
clusters at the lattice sites (Fig.~\ref{fig:topoext}(c)), and the cluster
centers are the reference positions. The twelve outer ones are dropped,
leaving the thirteen central points whose convex hull is the square drawn in
main Fig.~4. The black points in main Fig. 4 are the instantaneous topological points of the condition shown, extracted at regular time intervals throughout the record and overlaid.

\section{Supplemental movies}
\label{sisec:movies}
Four movies (MPEG-4, H.264) accompany this document. All show the
$20\times20~\mathrm{cm}^2$ interrogation area of the main text with axes in
cm, and each condition is labeled by $N$ and $E$ as in
Table~\ref{tab:conditions}.

\medskip\noindent\textbf{Movie S1} ($10~\mathrm{s}$, real time). Camera
images of two runs at $E\approx0.25~\mathrm{cm^2\,s^{-2}}$ with $N=66$
(left) and $N=414$ (right). Every second frame is shown at $30$ frames per
second. Top row: the recorded images after post-processing for display
(background subtraction and contrast normalization of the tracers). Middle
row: the images used as input to the particle tracking, in which the shrimp
have been removed by the detection step of Sec.~\ref{sisec:shrimp}, with
the same contrast enhancement. Bottom row: the shrimp regions removed from
the same frames. The small dots are the tracers and the larger objects are
the shrimp.

\medskip\noindent\textbf{Movie S2} ($20~\mathrm{s}$, $1.7\times$ real
time). The nine conditions of main Fig.~3 in the same arrangement: $100$
consecutive $\mathrm{FTLE}^{-}$ snapshots ($33~\mathrm{s}$ of record) with
the shrimp positions every fifth frame.

\medskip\noindent\textbf{Movie S3} ($15~\mathrm{s}$, $1.7\times$ real
time). Construction of the stability $R$ of the main text for the run with
$N=54$ and $E=0.18~\mathrm{cm^2\,s^{-2}}$ (top row, $R=0.63$) and for the
shrimp-free run at $E=1.21~\mathrm{cm^2\,s^{-2}}$ (bottom row, $R=0.12$).
The columns show the LCS region at time $t$, the region one turnover time
later ($T_e=8.4$ and $3.2~\mathrm{s}$ for the two runs), and their
overlap: grey where both contain the point, blue where only the earlier
one does, and red where only the later one does. The title of the third
column gives the Jaccard index of the pair shown and the run mean $R$.

\medskip\noindent\textbf{Movie S4} ($10~\mathrm{s}$, $3.3\times$ real
time). The nine conditions of main Fig.~4 in the same arrangement. Each
panel shows the time-smoothed curvature field $\overline{\log_{10}\kappa}$
of Sec.~\ref{sisec:topo}, the topological points of that frame (black), and the thirteen
reference points with their hull (red).

\bibliography{PRF}